\documentclass[aip,jcp,11pt,reprint,floatfix,superscriptaddress,urlname]{revtex4-1}

\usepackage[utf8]{inputenc}

\usepackage{booktabs}
\usepackage{amsmath}
\usepackage{color}
\usepackage{comment}
\usepackage{caption}
\usepackage{subcaption}
\usepackage{algorithm}
\usepackage{graphicx}
\usepackage{xcolor}

\usepackage[a4paper,right=15mm ,left=15mm,top=15mm,bottom=15mm]{geometry}

\newcommand{\bra}[1]{{\big< #1 \big|}}
\newcommand{\ket}[1]{{\big| #1 \big>}}

\DeclareMathAlphabet      {\mathbfit}{OML}{cmm}{b}{it}

\begin{document}

\title{Fully analytic G$_0$W$_0$ nuclear gradients}

\author{Johannes T\"olle}
\email{jojotoel@gmail.com}
\noaffiliation
\affiliation{Theoretische Organische Chemie, Organisch-Chemisches Institut and \\
Center for Multiscale Theory and Computation (CMTC),\\
Universität Münster, Corrensstraße 40, 48149 Münster, Germany}

\begin{abstract}
In this letter, we present the first fully analytic derivation and implementation of nuclear gradients for the G$_0$W$_0$ method. For this, we leverage the recently established connection between the G$_0$W$_0$ approach and equation-of-motion unitary coupled-cluster theory for charged excitations [J. Chem. Phys. 158, 124123 (2023)]. Analytic gradients are obtained through the Lagrangian technique and are implemented and validated by comparison with finite-difference calculations. For G$_0$W$_0$, we examine the effect of the Tamm--Dancoff approximation for evaluating the screened Coulomb interaction. Finally, we compare G$_0$W$_0$ adiabatic ionization potential and electron affinities to wavefunction-based electronic structure methods, and experimental values.
\end{abstract}
\maketitle 
\section{Introduction}
The GW method\cite{hedin1965new} is a well-established approach for determining quasiparticle (associated with electron removal and addition processes) energies in condensed matter physics, c.f.~Refs.~\citenum{hedin1970effects,aryasetiawan1998gw,hedin1999correlation,golze2019gw}. 
Particularly, the non-self-consistent G$_0$W$_0$ variant is widely used due to its balance between computational cost and accuracy \cite{bruneval2021gw}. 
Within the last decade, the GW (G$_0$W$_0$) method has also been adopted in molecular quantum chemistry with great success \cite{bruneval2012ionization,van2013gw,van2015gw,foerster2020low,kaplan2015off,wilhelm2018toward,golze2018core,forster2021low,forster2021gw100,bruneval2021gw,li2022benchmark,panades2023accelerating,marie2023gw}. 
In spite of its popularity, the determination of exact analytic properties, e.g., quasiparticle nuclear gradients, has not been achieved, yet. 
\\
Having access to these quantities is, however, essential for determining quantities such as adiabatic ionization potentials and electron affinities, which are of central importance for understanding redox processes in molecular and condensed matter systems, c.f.~Refs.\citenum{korth2014large,gaiduk2018electron,diddens2022modeling}.
Furthermore, the gradients allow for the determination of electron-phonon couplings that include many-body electron correlation effects. The necessity for including these effects has been highlighted, for example, in Refs.~\citenum{lazzeri2008impact,faber2015exploring,monserrat2016correlation,li2019electron}. 
To date, their analytic evaluation within the G$_0$W$_0$ approximation has been only achieved approximately, c.f.~Refs.~\citenum{lazzeri2008impact,faber2011electron,li2019electron,li2024electron}.
Furthermore, they allow for the determination of zero-point renormalization effects \cite{alvertis2024influence}, and many more important properties, c.f.~Refs.~\citenum{bernardi2016first,giustino2017electron}.
Lastly, G$_0$W$_0$ nuclear gradients constitute an important ingredient towards determining fully analytic G$_0$W$_0$-BSE \cite{onida2002electronic,blase2018bethe,blase2020bethe,bintrim2022full} nuclear gradients \cite{ismail2003excited,caylak2021machine,knysh2022modeling}.
\\
In this work, we demonstrate how G$_0$W$_0$ quasiparticle nuclear gradients can be obtained through the recent reformulation of G$_0$W$_0$ in Ref.~\citenum{tolle2023exact}, together with the Lagrangian technique \cite{helgaker1982simple,helgaker1989numerically,koch1990coupled}.
\begin{figure}
    \includegraphics[width=0.5\textwidth]{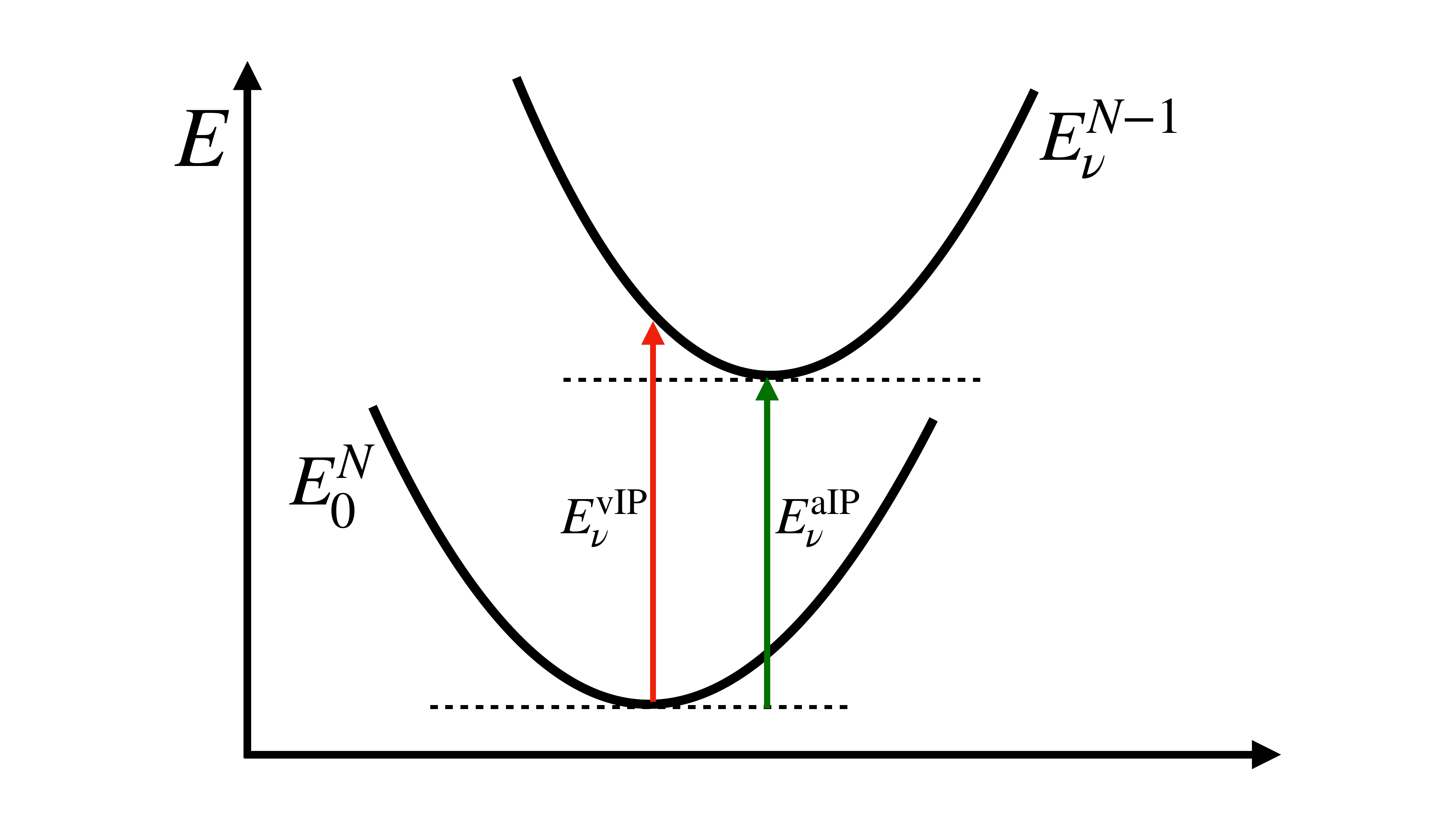}
    \caption{Illustration of vertical and adiabatic ionization potential (vIP/aIP) of state $\nu$.}
    \label{fig:Illustration}
\end{figure}
Another prerequisite, when determining relaxed excited state nuclear gradients, arises from identifying the proper ground-state energy $E^N_0$ (compare Fig.~\ref{fig:Illustration}).
From Ref.~\citenum{tolle2023exact}, $E^N_0$ can be unambiguously identified as the direct Random-Phase-Approximation (dRPA) [or direct-ring unitary coupled-cluster doubles (drUCCD)] energy \cite{tolle2023exact}. 
Analytic gradients for drUCCD will therefore also be presented, which constitute an alternative route for determining dRPA nuclear gradients \cite{burow2014analytical,ramberger2017analytic,stein2024massively}.
Lastly, the effect of the Tamm--Dancoff Approximation (TDA), for which also $E^N_0$ reduces to the mean-field energy, for evaluating the screened Coulomb interaction will be investigated. 
\section{Theory}
Within the G$_0$W$_0$ approximation, the $\nu$-th quasiparticle energy ($E_\nu^\mathrm{QP}$) is obtained as
\begin{align}
    \left[ \mathbf{F} + \pmb{\Sigma}_\mathrm{c}(E_\nu^\mathrm{QP})\right] \mathbf{U}_\nu = E_\nu^\mathrm{QP}\mathbf{U}_\nu,
\end{align}
where $\mathbf{F}$ denotes the Fock matrix (in the Hartree--Fock approximation), $\pmb{\Sigma}_\mathrm{c}(E_\nu^\mathrm{QP})$ the energy dependent correlation self-energy, and $\mathbf{U}_\nu$ the excitation vector.
In Ref.~\citenum{tolle2023exact} the exact relationship of this procedure to an equation-of-motion (EOM) coupled-cluster (CC) approach for ionization potentials (IP) and electron affinities (EA) within the quasi-boson formalism has been demonstrated.
\\
Within the quasi-boson formalism, the electron-boson linear coupling Hamiltonian, underlying the G$_0$W$_0$ approximation, is \cite{tolle2023exact,tolle2024ab}
\begin{align}
    \hat{H} &= \hat{F}_\mathrm{e} + \hat{V}_\mathrm{eB} + \hat{H}_\mathrm{B}  \\
    \hat{F}_\mathrm{e} &= \sum_{pq} F_{pq} \left\{\hat{a}^\dagger_p \hat{a}_q\right\}\\
    \hat{V}_\mathrm{eB} &= \sum_{pq,I} V^I_{pq} \left\{\hat{a}^\dagger_p \hat{a}_q\right\} \left(\hat{b}^\dagger_I + \hat{b}_I \right)\\
    \hat{H}_\mathrm{B} &= \sum_{IJ} A_{IJ} \hat{b}^\dagger_I \hat{b}_J + \frac{1}{2} \sum_{IJ} B_{IJ} \left( \hat{b}^\dagger_I \hat{b}^\dagger_J + \hat{b}_I \hat{b}_J\right),
\end{align}
where $F_{pq}$ denotes the electronic Fock matrix, $\hat{a}^\dagger_p$($\hat{b}^\dagger_I$)/$\hat{a}_p$($\hat{b}_I$) denote Fermionic (Bosonic) creation/annihilation operators, and $\{ \dots \}$ normal-ordered fermionic operators with respect to the Fermi vacuum. $V^I_{pq}$ is
\begin{align}
    V^I_{pq} = (pq|I) = (pq|ia) = V^{ia}_{pq},
\end{align}
\begin{align}
    A_{IJ} = A_{ia,jb} = \delta_{ij}  F_{ab} - \delta_{ab} F_{ij} + (ia|jb),
    \label{eq:AMatrix}
\end{align}
and 
\begin{align}
    B_{IJ} = B_{ia,jb} = (ia|jb).
\end{align}
In the following, the usual notation of electronic indices ($i,j,k,\dots$ denote occupied orbitals, $a,b,c,\dots$ denote virtual orbitals, and $p,q,r,s,\dots$ general orbital indices) is used, and electron repulsion integrals are expressed in $(11|22)$ (Mulliken) notation. 
Furthermore, real orbitals are assumed throughout.
Bosonic indices are denoted with capital Latin letters $I,J,K,\dots$ (referring to composite Fermionic indices $I=ia$). 
All non-number conserving Bosonic contributions can be removed by unitary transformation
\begin{align}
    \bar{H} = e^{-\hat{\sigma}} \hat{H} e^{\hat{\sigma}},
\end{align}
with
\begin{align}
    \hat{\sigma} = \frac{1}{2} \sum_{IJ} t_{IJ} \left( \hat{b}^\dagger_I \hat{b}^\dagger_J - \hat{b}_I \hat{b}_J \right).
\end{align}
The amplitudes $t_{IJ}$ are determined from
\begin{align}
    J_{IJ} = \bra{0} \left[\bar{H},\hat{b}^\dagger_I \hat{b}^\dagger_J - \hat{b}_I \hat{b}_J \right] \ket{0} \overset{!}{=} 0,
    \label{eq:AmplitudeEquation}
\end{align}
and the ground-state [direct Random-Phase Approximation (dRPA)] energy is \cite{ring2004nuclear} 
\begin{align}
    E^\mathrm{dRPA} = \bra{0} \bar{H} \ket{0} + E^\mathrm{HF}.
    \label{eq:dRPAEnergy}
\end{align}
Note that the above procedure is identical to the direct-ring unitary coupled-cluster doubles (drUCCD) method, i.e.~$E^\mathrm{dRPA}=E^\mathrm{drUCC}$. Furthermore, we assume a Hartree--Fock (HF) reference mean-field state throughout for simplicity. 
A generalization to other mean-field references is straightforward.
Additionally, $\ket{0}$ refers to $\ket{0_\mathrm{F} 0_\mathrm{B}}$, i.e, the combined Fermionic and Bosonic reference state. 
\\
Within the quasi-boson formulation of G$_0$W$_0$, quasiparticle energies (associated with $N\pm1$ processes) are obtained through the equation-of-motion (EOM) approach \cite{tolle2023exact,tolle2024ab}.  
In this case the excitation operator $\hat{R}^\mathrm{QP}_\nu$ for quasiparticle $\nu$ reads
\begin{align}
    \hat{R}^\mathrm{QP}_\nu &= \sum_m \hat{C}_{m} R_{\nu,m} \nonumber \\
    &= \sum_i \hat{a}_i R_{\nu,i}  + \sum_a  \hat{a}_a R_{\nu,a} \nonumber \\
    &+ \sum_{iI} \hat{a}_i \hat{b}^\dagger_I R_{\nu,iI}  + \sum_{Ia}  \hat{b}_I \hat{a}_a R_{\nu,Ia},
\end{align}
and quasiparticle energies are obtained from
\begin{align} 
    \tilde{\mathbf{H}}^\mathrm{QP} \mathbf{R}_\nu = E^\mathrm{QP}_\nu \mathbf{R}_\nu .
\end{align}
$\tilde{\mathbf{H}}^\mathrm{QP}$ denotes
\begin{align}
    \tilde{\mathbf{H}}^\mathrm{QP} = \mathbf{S}^{-1} \mathbf{H}^\mathrm{QP},
\end{align}
where 
\begin{align}
    H^\mathrm{QP}_{nm} = \bra{0} \left[ \hat{C}^\dagger_n \left[ \bar{H}, \hat{C}_m \right] \right] \ket{0},
    \label{eq:QPHamiltonian}
\end{align}
and 
\begin{align}
    S_{nm} = \bra{0} \left[  \hat{C}^\dagger_n, \hat{C}_m \right] \ket{0}.
\end{align}
Note that in this convention the quasiparticles are associated with the following physical processes depending on whether they describe holes or particles
\begin{align}
    \textrm{hole:}~&E^\mathrm{QP}_\nu = E^{N}_0-E^{N-1}_\nu, \nonumber \\
    \textrm{particle:}~&E^\mathrm{QP}_\nu = E^{N+1}_\nu-E^{N}_0.
    \label{eq:PHDefinition}
\end{align}
Here response properties of the G$_0$W$_0$ method, e.g., nuclear gradients, are determined by the Lagrangian technique. 
The Lagrangian $L^\mathrm{QP}_\nu$ for quasiparticle $\nu$ is
\begin{align}
     L^{\mathrm{QP}}_\nu(\mathbf{R}^\mathrm{QP}_\nu, \pmb{\lambda}, \mathbf{M}, \mathbf{Z}, \mathbf{C}, \mathbf{t}) &= f_\nu G^\mathrm{QP}_{\nu}[\mathbf{R}^\mathrm{QP}_\nu,E^\mathrm{QP}_\nu] + E^\mathrm{drUCCD} \nonumber \\
     &+ \sum_{IJ} Z_{IJ} J_{IJ} \nonumber \\
     &+ \sum_{ia} \lambda_{ia} F_{ia} \nonumber \\
     &+ \sum_{pq} M_{pq} \left( S_{pq} - \delta_{pq} \right),
     \label{eq:QPLagrangian}
\end{align}
with $Z_{IJ}$, $\lambda_{ia}$, and $M_{pq}$ denoting the Lagrange multipliers.
In particular, $Z_{IJ}$ enforces the amplitude equation [Eq.~(\ref{eq:AmplitudeEquation})], $\lambda_{ia}$ the Brillouin condition, and $M_{pq}$ the orthonormality of the molecular orbitals. 
Note that atom-centered basis functions are assumed throughout.
$f_\nu$ denotes the sign function, i.e.,
\begin{align}
    f_\nu = \mathrm{sgn} \left( E^\mathrm{QP}_\nu - \epsilon_\mathrm{F}\right), 
\end{align}
and $\epsilon_\mathrm{F}$ the Fermi energy.
$f_\nu$ is used to obtain total $E^{N+1}_\nu$/$E^{N-1}_\nu$ energies [compare Eq.~(\ref{eq:PHDefinition})].  
The quasiparticle functional $G^\mathrm{QP}_{\nu}[\mathbf{R}_\nu,E^\mathrm{QP}_\nu]$ of excitation $\nu$ is \cite{furc2002}
\begin{align}
    G^\mathrm{QP}_{\nu}[\mathbf{R}^\mathrm{QP}_\nu,E^\mathrm{QP}_\nu] =& \mathbf{R}^{\mathrm{QP},T}_\nu \tilde{\mathbf{H}}^\mathrm{QP} \mathbf{R}^\mathrm{QP}_\nu \nonumber \\
    &- E^\mathrm{QP}_\nu \left( \mathbf{R}^{\mathrm{QP},T}_\nu \mathbf{R}^{\mathrm{QP}}_\nu - 1\right).
\end{align} 
After making the Lagrangian stationary with respect to its parameters $\mathbf{R}^\mathrm{QP}_\nu,\pmb{\lambda}$, $\mathbf{M}$, $\mathbf{C}$ (orbital coefficients), $\mathbf{Z}$, and $\mathbf{t}$, the nuclear gradients becomes: $\mathbf{G}^\mathrm{QP}_{\nu} = \frac{\partial L^\mathrm{QP}_\nu}{\partial \mathbf{R}}$.
Note that gradients within the Tamm--Dancoff Approximation (TDA) for the screened Coulomb interaction (G$_0$W$_0$-TDA) are obtained by substituting $\bar{H}$ with $\hat{H}$ in Eq.~\ref{eq:QPHamiltonian}.
In this case, neither the amplitudes $\mathbf{t}$ nor the multipliers $\mathbf{Z}$ need to be considered in $L^{\mathrm{QP}}_\nu$, resulting in a lower computational scaling of $\mathcal{O}(N^5)$ compared to G$_0$W$_0$ [$\mathcal{O}(N^6)$]. 
\\
Details regarding EOM-CC nuclear gradients in general are presented in Refs.~\citenum{stanton1993many,stanton1994analytic,levchenko2005analytic}.
Specific working equations and implementation details for G$_0$W$_0$ are given in the Appendix (Sec.~\ref{sec:Appendix}). 

\section{Computational details}
A first pilot implementation of the G$_0$W$_0$ nuclear gradients has been realized in a python code that heavily relies on functionalities provided by $\textsc{PySCF}$ \cite{sun2018pyscf,sun2020recent}.
Throughout this manuscript, no diagonal approximation for determining quasiparticle energies is imposed. 
Geometry optimizations have been performed using the \textsc{PyBerny} \cite{pyberny} interface of \textsc{PySCF}, and the following modified thresholds are used: maximum component and root mean square of the gradient (`gradientmax'/ `gradientrms') are set to  $10^{-4}$ Ha/Bohr, and the largest value and root mean square in the change of coordinates (`stepmax'/ `steprms') are set to $10^{-4}$ Bohr.
\\
The drUCCD amplitudes are determined following the procedure described in Ref.~\citenum{tolle2024ab}. At convergence, the change of the amplitudes from the previous iteration is $<10^{-10}$. Quasiparticle energies have been determined within the EOM approach \cite{bintrim2021full,tolle2023exact} using the Davidson procedure until the change in the quasiparticle energy is smaller than $10^{-10}$ Ha. 
Additional technical thresholds are presented in the Appendix (Sec.~\ref{sec:Appendix}).
\\
SCF and geometry optimizations for systems with an odd number of electrons have been performed based on the restricted open-shell framework, as implemented in \textsc{PySCF}.
\section{Results}
\subsection{Finite differences}
In the first step, the completeness of the Lagrangian [Eq.~\ref{eq:QPLagrangian}], and the correctness of the implementation is demonstrated by comparing the analytic evaluation of the nuclear gradients to numerical nuclear gradients obtained from finite difference calculations. In the latter case a step-size of $\Delta = 0.01$ Ha/\AA~(along the bond-direction) in a four-point formula has been used
\begin{align}
    G_\mathrm{num} &=  \frac{1}{12\Delta} \left( E[R_b - 2 \Delta ] - 8E[R_b - \Delta ] \right. \nonumber \\
    &\left. + 8E[R_b + \Delta ] - E[R_b + 2\Delta ] \right). 
\end{align}  
The absolute deviation of the gradient for drUCCD, G$_0$W$_0$ and  G$_0$W$_0$-TDA are presented in Tab.~\ref{tab:druCCDGradTest}, Tab.~\ref{tab:EOMIPGRAD}, and Tab.~\ref{tab:EOMEAGRAD} respectively.
It can be seen that the gradients agree within $10^{-6}$ Ha/\AA. 
\begin{table}[h]
    \centering
    \caption{drUCCD numerical $G_\mathrm{num}$, analytic $G_\mathrm{an}$ nuclear gradient along the bond direction, and absolute difference $\Delta G$ for five diatomic molecules in Ha/\AA~(HF/cc-pvtz). }
    \begin{tabular}{l c c c}
        \hline
         Molecule & $G_\mathrm{num}$ & $G_\mathrm{an}$ & $\Delta G$\\
         \hline
         H$_2$$^{a}$	&0.008526&	0.008526&	0.000000\\
         HCl$^{b}$ &0.010702&	0.010702&	0.000000\\
         HF$^{b}$	&0.020717&	0.020716&	0.000001\\
         N$_2$$^{a}$	&0.079904&	0.079904&	0.000000\\
         CO$^{a}$	&0.052797&	0.052797&	0.000000\\
         \hline
    \multicolumn{4}{c}{
    \begin{footnotesize}
        a) Bond length taken from Ref.~\citenum{dalskov1998correlated}
    \end{footnotesize}
    } \\
    \multicolumn{4}{c}{
    \begin{footnotesize}
        b) taken from Ref.~\citenum{sekino1993molecular}
    \end{footnotesize}
    }
    \end{tabular}
    \label{tab:druCCDGradTest}
\end{table}
\begin{table*}[t]
    \centering
    \caption{$E^{N-1}$ (with and without the TDA) numerical $G_\mathrm{num}$, analytic $G_\mathrm{an}$ nuclear gradients along the bond direction, and absolute difference $\Delta G$ for five diatomic molecules in Ha/\AA~(HF/ccpvtz). }
    \begin{tabular}{l c c c c c c}
    \hline
                  & \multicolumn{3}{c }{G$_0$W$_0$-TDA} & \multicolumn{3}{ c}{G$_0$W$_0$}  \\
        \hline
         Molecule & $G_\mathrm{num}$ & $G_\mathrm{an}$ & $\Delta G$ & $G_\mathrm{num}$ & $G_\mathrm{an}$ & $\Delta G$\\
         \hline
         H$_2$$^{a}$&-0.261017&	-0.261017&	0.000000	&-0.270697	&-0.270698&	0.000001\\
         HCl$^{b}$	&-0.034377&	-0.034377&	0.000000	&-0.027896	&-0.027896&	0.000000 \\ 
         HF$^{b}$	&-0.108402&	-0.108402&	0.000000	&-0.116077	&-0.116078&	0.000001 \\
         N$_2$$^{a}$&-0.118181&	-0.118181&	0.000000	&-0.274598	&-0.274598&	0.000000 \\ 
         CO$^{b}$	&0.219044 &	 0.219044&	0.000000	&0.157933	&0.157932 & 0.000001 \\
    \hline
    \multicolumn{5}{c}{
    \begin{footnotesize}
        a) Bond length taken from Ref.~\citenum{dalskov1998correlated},
        b) taken from Ref.~\citenum{sekino1993molecular}
    \end{footnotesize}
    }
    \end{tabular}
    \label{tab:EOMIPGRAD}
\end{table*}
\begin{table*}[t]
    \centering
    \caption{$E^{N+1}$ (with and without the TDA) numerical $G_\mathrm{num}$, analytic $G_\mathrm{an}$ nuclear gradients along the bond direction, and absolute difference $\Delta G$ for five diatomic molecules in Ha/\AA~(HF/ccpvtz). }
    \begin{tabular}{l c c c c c c}
    \hline
        & \multicolumn{3}{c}{G$_0$W$_0$-TDA} & \multicolumn{3}{c}{G$_0$W$_0$}  \\
        \hline
         Molecule & $G_\mathrm{num}$ & $G_\mathrm{an}$ & $\Delta G$ & $G_\mathrm{num}$ & $G_\mathrm{an}$ & $\Delta G$\\
         \hline
         H$_2$$^{a}$&-0.145433&	-0.145433&	0.000000	&-0.145875&	-0.145875&	0.000000 \\
         HCl$^{b}$	&-0.203715&	-0.203715&	0.000000	&-0.200577&	-0.200578&	0.000001 \\
         HF$^{b}$	&-0.132585&	-0.132586&	0.000001	&-0.158019&	-0.158019&	0.000000 \\ 
         N$_2$$^{a}$&-0.287638&	-0.287638&	0.000000	&-0.416743&	-0.416744&	0.000001 \\
         CO$^{b}$	&-0.262626&	-0.262626&	0.000000	&-0.334561&	-0.334561&	0.000000 \\
    \hline
    \multicolumn{5}{c}{
    \begin{footnotesize}
        a) Bond length taken from Ref.~\citenum{dalskov1998correlated},
        b) taken from Ref.~\citenum{sekino1993molecular}
    \end{footnotesize}
    }
    \end{tabular}
    \label{tab:EOMEAGRAD}
\end{table*}
\subsection{Excited state relaxation}
\begin{figure}[b]
    \includegraphics[width=0.45\textwidth]{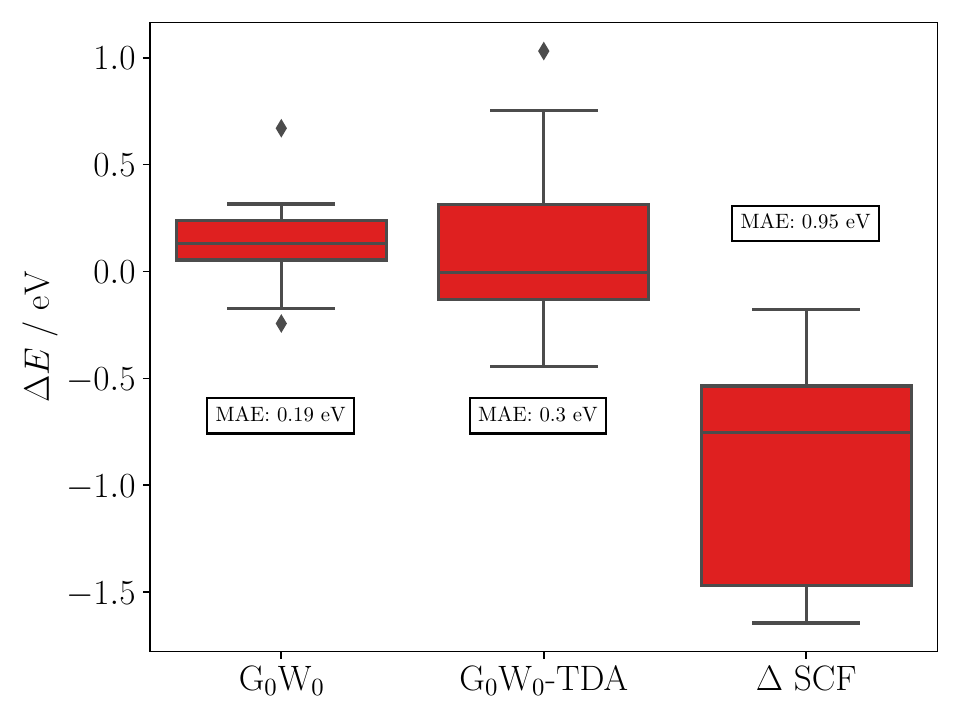}
    \caption{Error distribution and mean-absolute error (MAE) for G$_0$W$_0$, G$_0$W$_0$-TDA, and $\Delta$SCF within the Hartree--Fock approximation relative to experimental values for 15 processes (displayed in Tab.~\ref{tab:Processes}) in eV. The aug-cc-pvtz basis set is used throughout.}
    \label{fig:ExpTheoryBox}
\end{figure}

Having established the overall correctness of the nuclear gradients, they can be used for determining bond lengths and adiabatic energies of electron-attaching (aEA)/ionizing processes (aIP), as illustrated in Fig.~\ref{fig:Illustration}. 
\\
First, the bond length of different (energetically lowest) ionization and electron addition processes in diatomic molecules is investigated.
The results are presented in Tab.~\ref{tab:Bonddiatomics} together with alternative wavefunction-based quantum chemistry procedures (ADC(3) and EOM-CCSD taken from Ref.~\citenum{rehn2024analytical}).
In all cases, smaller bond length for G$_0$W$_0$-TDA compared to G$_0$W$_0$ and a larger deviation from the experimental value for G$_0$W$_0$-TDA compared to G$_0$W$_0$ is observed. 
Furthermore, a good agreement between G$_0$W$_0$, ADC(3)/CCSD, and experiment is found. 
\begin{table}[h!]
    \centering
    \caption{Bond-length of different electron addition and removal processes for diatomic molecules in~\AA~(HF/aug-cc-pvqz). CCSD refers to EOM-EA/IP-CCSD,}
    \begin{tabular}{l r r r r r r}
        \hline
        Process                  & G$_0$W$_0$ & G$_0$W$_0$-TDA & ADC(3)$^a$ & CCSD$^a$ & experiment$^b$ \\
        \hline
        NO$^+$$\rightarrow$NO$^{\cdot}$ &    1.139        &      1.100          &   1.123 & 1.138& 1.154 \\
        OH$^{-}$$\rightarrow$OH$^{\cdot}$ &    0.951        &     0.947         &  0.951 & 0.966 & 0.970 \\
        CO$\rightarrow$CO$^+$ & 1.092  &  1.081  & 1.108 & 1.104 & 1.128 \\
        \hline 
        \multicolumn{5}{c}{
            \begin{footnotesize}
                a) taken from Ref.~\citenum{rehn2024analytical}
            \end{footnotesize}
        }&&\\
        \multicolumn{5}{c}{
            \begin{footnotesize}
                b) taken from Ref.~\citenum{cccbdb}
            \end{footnotesize}
        }&&\\
    \end{tabular}
    \label{tab:Bonddiatomics}
\end{table}
\\
Next, the lowest aIPs and aEAs for G$_0$W$_0$, G$_0$W$_0$-TDA, and $\Delta$SCF (Hartree--Fock) versus experimental estimates (taken from~\citenum{curtiss1991gaussian}) for 15 electron addition and removal processes in small molecular systems are investigated. 
The results are shown in Fig.~\ref{fig:ExpTheory}, and the individual processes together with their experimental values are listed in the Appendix, Tab.~\ref{tab:Processes}.
The error distributions and the mean-absolute errors (MAE) are displayed in Fig.~\ref{fig:ExpTheoryBox}.
\begin{figure}[t]
    \includegraphics[width=0.45\textwidth]{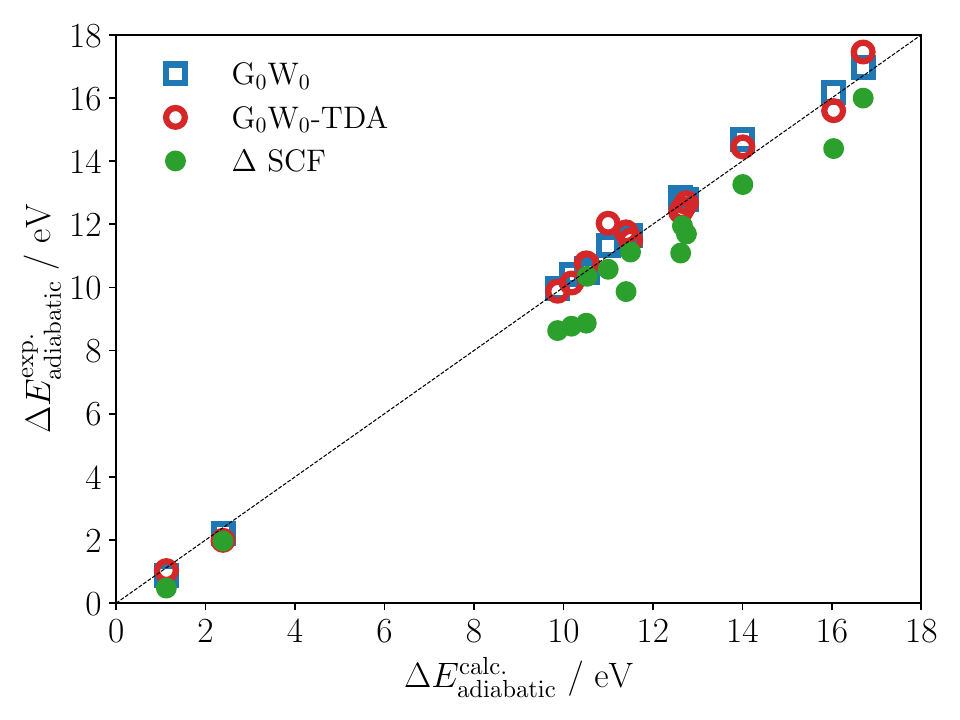}
    \caption{Experimental versus theoretical (G$_0$W$_0$, G$_0$W$_0$-TDA, and $\Delta$ SCF within the Hartree--Fock approximation) adiabatic ionization potentials and electron affinities for 15 processes (displayed in Tab.~\ref{tab:Processes}) in eV. The aug-cc-pvtz basis set is used throughout.}
    \label{fig:ExpTheory}
\end{figure}
The best agreement with experimentally determined energies is observed for G$_0$W$_0$ (MAE $=0.19$ eV) with an accuracy comparable to the performance of the G$_0$W$_0$ method for determining vertical quasiparticle energies \cite{marom2012benchmark,bruneval2021gw,forster2021gw100,golze2022accurate,marie2023gw}.
Approximating the screened Coulomb interaction within the TDA results in an MAE $=0.30$ eV. 
The largest deviation (MAE $=0.95$ eV) is obtained for adiabatic energies determined from the $\Delta$ SCF procedure.
\\
Next, the aIPs and aEAs for various medium-sized organic molecules in the cc-pvdz basis set are compared. 
The aIPS for G$_0$W$_0$ and G$_0$W$_0$-TDA are shown in Tab.~\ref{tab:AdiabaticIP}.
Tab.~\ref{tab:AdiabaticIP} also contains aIPs for ADC(2) and EOM-IP-CCSD calculations, taken from Ref.~\citenum{rehn2024analytical}.
The MAE for both G$_0$W$_0$ and G$_0$W$_0$-TDA with respect to experiment is $0.19$.
The largest G$_0$W$_0$ deviation of $0.36$ eV is found for quinone while the largest deviation for G$_0$W$_0$-TDA is $-0.36$ eV for $\alpha$-pyranose.
For quinone, nitrobenzene, and uracil, both G$_0$W$_0$ and G$_0$W$_0$-TDA, are considerably more accurate than ADC(2), relative to EOM-IP-CCSD and experiment. 
The aEAs are displayed in Tab.~\ref{tab:AdiabaticEA}.
There it can be seen that the overall deviation relative to the experimental aEAs is increased, resulting in an MAE of 0.97~eV/0.93~eV for G$_0$W$_0$ and G$_0$W$_0$-TDA, respectively.
However, for quinone, nitrobenzene, and uracil an excellent agreement (largest deviation of 0.15~eV for G$_0$W$_0$-TDA) with aEAs from EOM-EA-CCSD\cite{rehn2024analytical} is found. In contrast, ADC(2)\cite{rehn2024analytical} results in substantially larger deviations relative to EOM-EA-CCSD.
Disentangling the precise origin of the deviations from the experiment, both for G$_0$W$_0$ and G$_0$W$_0$-TDA, would require the evaluation of aIPs and aEAs in considerably larger basis sets, and the consideration of zero point vibrational energy corrections, which is outside the scope of this work.
\begin{table}[t]
    \centering
    \caption{Adiabatic ionization potential (aIP) for medium-sized organic molecules for various electronic structure methods in eV (HF/cc-pvdz). The mean-absolute error (MAE) for G$_0$W$_0$ and G$_0$W$_0$-TDA with respect to experimental values are also shown. CCSD refers to EOM-IP-CCSD}
    \begin{tabular}{l r r r r r}
        \hline
        Molecule & G$_0$W$_0$ & G$_0$W$_0$-TDA & ADC(2)$^\mathrm{a}$ & CCSD$^\mathrm{a}$ & experiment \\
        \hline
        quinone           & 10.36 & 9.74  & 7.61 & 9.75  & 10.00$\pm0.1$$^{b}$\\
        nitrobenzene      & 9.81  & 9.88  & 9.18 & 9.73  & 9.94$\pm0.08$$^{b}$\\
        uracil            & 9.24  & 9.17  & 7.39 & 8.92  & 9.20$^{b}$\\
        thymine           & 8.83  & 8.78  &  -   &  -    & 8.95$\pm0.05$$^{c}$\\
        benzene           & 8.95 & 9.01   & -    &  -    &9.24$^{b}$\\
        naphthalene        & 7.81& 7.93    & -    &  -    & 8.14$^{b}$\\
        $\alpha$-pyranose & 9.03 & 8.74   & -    &  -    & 9.10$^d$\\
        \hline 
        MAE               & 0.19  & 0.19  & -  &  - &  - \\
        \hline 
        \multicolumn{3}{c}{
            \begin{footnotesize}
                a) taken from Ref.~\citenum{rehn2024analytical}
            \end{footnotesize}
        }&&&\\
        \multicolumn{3}{c}{
            \begin{footnotesize}
               b) taken from Ref.~\citenum{nist_asd}
            \end{footnotesize}
        }&&&\\
        \multicolumn{3}{c}{
            \begin{footnotesize}
               c) taken from Ref.~\citenum{bravaya2010electronic}
            \end{footnotesize}
        }&&&\\
        \multicolumn{3}{c}{
            \begin{footnotesize}
               d) taken from Ref.~\citenum{ghosh2012vuv}
            \end{footnotesize}
        }&&&\\
    \end{tabular}
    \label{tab:AdiabaticIP}
\end{table}

\begin{table}[t]
    \centering
    \caption{Adiabatic electron affinities (aEAs) for medium-sized organic molecules for various electronic structure methods in eV (HF/cc-pvdz). The mean-absolute error (MAE) for G$_0$W$_0$ and G$_0$W$_0$-TDA with respect to experimental values are also shown. CCSD refers to EOM-EA-CCSD.}
    \begin{tabular}{l r r r r r}
        \hline
        Molecule & G$_0$W$_0$ & G$_0$W$_0$-TDA & ADC(2)$^\mathrm{a}$ & CCSD$^\mathrm{a}$ & experiment \\
        \hline
        quinone           &    1.00 &    0.90 & 2.21 & 1.05  & 1.85$\pm0.01$$^{b}$\\
        nitrobenzene      &    0.02 & $-$0.03 & 1.17 & 0.00  & 1.00$\pm0.01$$^{b}$\\
        uracil            & $-$0.98 & $-$1.01 & 0.29 & $-$0.94 & 0.08$^{b}$\\
        benzene           & $-$2.27 & $-$2.06 & -    &  -    & $-1.15$$^{c}$\\
        naphthalene       & $-$1.04 & $-$0.88 & -    &  -    & $-0.20$$^{b}$\\
        \hline 
        MAE               & 0.97  &  0.93 & -  &  - &  - \\
        \hline 
        \multicolumn{3}{c}{
            \begin{footnotesize}
                a) taken from Ref.~\citenum{rehn2024analytical}
            \end{footnotesize}
        }&&&\\
        \multicolumn{3}{c}{
            \begin{footnotesize}
               b) taken from Ref.~\citenum{nist_asd}
            \end{footnotesize}
        }&&&\\
        \multicolumn{3}{c}{
            \begin{footnotesize}
               c) taken from Ref.~\citenum{jordan1976electron}
            \end{footnotesize}
        }&&&\\
    \end{tabular}
    \label{tab:AdiabaticEA}
\end{table}

\section{Conclusion and outlook}
In summary, fully analytic nuclear gradients for the G$_0$W$_0$ approximation have been derived and implemented for the first time. 
The derivation is based on the recently established connection between the G$_0$W$_0$ method and EOM-drUCCD \cite{tolle2023exact} and highlights the importance of establishing connections between the fields of quantum-chemistry and theoretical condensed matter physics, see also Refs.~\citenum{scuseria2008ground,scuseria2013particle,lange2018relation,berkelbach2018communication,quintero2022connections}.
\\
Using these gradients, we have benchmarked the accuracy of the G$_0$W$_0$ method, with and without the TDA approximation for the screened Coulomb interaction, to determine the adiabatic electron affinities (aEA) and ionization potentials (aIP) for various molecular systems.
Overall, a similar accuracy for aEAs/aIPs as for vertical IPs/EAs is observed. 
This indicates that the `GW miracle of many-body perturbation theory \cite{bruneval2021gw} potentially also holds for linear response properties making it a viable alternative to wavefunction-based quantum chemical methods for molecular systems.
A more comprehensive assessment of aIPs/aEAs, and also other linear-response properties is planned in a follow-up study.
\\
Currently, the computational cost for evaluating G$_0$W$_0$ properties scales as $\mathcal{O}(N^6)$, and as $\mathcal{O}(N^5)$ for G$_0$W$_0$-TDA.
Future efforts aim to reduce this scaling, for instance, by employing the auxiliary-boson expansion introduced in Ref.~\citenum{tolle2024ab}.
Additionally, the approach will be extended to translationally invariant systems, enabling, among other advances, the determination of fully analytic electron-phonon couplings within the G$_0$W$_0$ approximation\cite{lazzeri2008impact,faber2011electron,li2019electron,li2024electron}. Finally, the derivation of analytic quasiparticle energy derivatives presented in this work marks an important step toward the development of fully analytic nuclear gradients within the G$_0$W$_0$-BSE framework.
\section*{Acknowledgement}
J.~T.~acknowledges fruitful discussions with Johannes~Neugebauer, and Anton~Rikus. 
Additionally, J.~T.~thanks  Johannes~Neugebauer, Pierre-Fran\c cois~Loos,  Thorsten~Deilmann, and Anton~Rikus for useful feedback on the manuscript.

\setcounter{equation}{0}\renewcommand\theequation{A\arabic{equation}}
\section{Appendix A: G$_0$W$_0$ nuclear gradients}
\label{sec:Appendix}
In practice, the nuclear gradients for quasiparticle $\nu$ are computed, after the Lagrangian is stationary with respect to its parameters, as 
\begin{align}
    \mathbf{G}^\mathrm{QP}_\nu &= \frac{\partial L^\mathrm{QP}_\nu(\mathbf{Z}, \pmb{\lambda}, \mathbf{M}, \mathbf{C}, \mathbf{t}) }{\partial \mathbf{R}} \nonumber \\
    &=  \sum_{pq} \frac{\partial h_{pq}}{\partial \mathbf{R}} \tilde{\gamma}_{pq}  + \frac{1}{2} \sum_{pqrs} \frac{\partial (pq|rs)}{\partial \mathbf{R}} \tilde{\Gamma}_{pqrs} \nonumber \\
    &+ \sum_{pq} \frac{\partial S_{pq}}{\partial \mathbf{R}} M_{pq}.
    \label{eq:EnergyGrad}
\end{align}
The one- and two-particle reduced density matrices $\tilde{\gamma}_{pq}$ and $\tilde{\Gamma}_{pqrs}$ are 
\begin{align}
    \tilde{\gamma}_{\nu,pq} = \gamma^\mathrm{MF}_{pq} + \gamma^\mathrm{drUCCD}_{pq} + \gamma^\mathrm{QP}_{\nu,pq} + \gamma^{\lambda}_{pq} + \gamma^\mathrm{Z}_{\nu,pq},
\end{align}
and 
\begin{align}
    \tilde{\Gamma}_{\nu,pqrs} &= \Gamma^\mathrm{drUCCD}_{pqrs} + \Gamma^\mathrm{QP}_{\nu,pqrs} + \Gamma^\mathrm{Z}_{pqrs} + \Gamma^{1}_{\nu,pqrs},
\end{align}
where $\Gamma^{1}_{\nu,pqrs}$ contains contributions from $\tilde{\gamma}_{\nu,pq}$ to $\tilde{\Gamma}_{\nu,pqrs}$ because of the definition of the energy functional [compare Eq.~(\ref{eq:EnergyGrad})].
$\gamma^\mathrm{MF}_{pq}$ denotes the mean-field density matrix, and $\gamma^\lambda_{pq}$ is the $\lambda$-Lagrange multiplier density matrix.
In case of the G$_0$W$_0$ nuclear gradients, $\gamma^\mathrm{drUCCD}_{pq}/\Gamma^\mathrm{drUCCD}_{pqrs},\gamma^\mathrm{QP}_{\nu,pq}/\Gamma^\mathrm{QP}_{\nu,pqrs}$ and $\gamma^\mathrm{Z}_{\nu,pq}/\Gamma^\mathrm{Z}_{pqrs}$ are method-specific and their derivation is outlined below.
\\
The drUCCD density matrices $\gamma^\mathrm{drUCCD}_{pq}/\Gamma^\mathrm{drUCCD}_{pqrs}$ are obtained as 
\begin{align}
    \gamma^\mathrm{drUCCD}_{pq} = \frac{\partial E^\mathrm{drUCCD}_\mathrm{c} }{\partial F_{pq}},
\end{align} 
and 
\begin{align}
    \Gamma^\mathrm{drUCCD}_{pqrs} = \frac{\partial E^\mathrm{drUCCD}_\mathrm{c} }{\partial V_{pqrs}},
\end{align}
where $E^\mathrm{drUCCD}_\mathrm{c}$ denotes the drUCCD correlation energy.
The contributions to $\gamma^\mathrm{drUCCD}_{pq}$ are
\begin{align}
    \gamma^\mathrm{drUCCD}_{ab} &= \sum_{i}\Gamma^A_{iaib} \nonumber \\
    \gamma^\mathrm{drUCCD}_{ij} &= \sum_{a} \Gamma^A_{iaja}.
\end{align}
Similarly for $\Gamma^\mathrm{drUCCD}_{pqrs}$, one fnds 
\begin{align}
    \Gamma^\mathrm{drUCCD}_{iajb} &= 2 \left( \Gamma^B_{iajb} +  \Gamma^A_{iajb} \right)
\end{align}
where we have made use of the fact that Bosonic indices correspond to Fermionic electron-hole pairs ($I\equiv ia$).
$\Gamma^A_{iajb}$ is $\bra{0} e^{-\hat{\sigma}}  \hat{b}^\dagger_{ia} \hat{b}_{jb} e^{\hat{\sigma}} \ket{0}$, and $\Gamma^B_{iajb} = \bra{0} e^{-\hat{\sigma}} \frac{1}{2} \left( \hat{b}^\dagger_{ia} \hat{b}^\dagger_{jb} + \hat{b}_{ia} \hat{b}_{jb} \right) e^{\hat{\sigma}} \ket{0}$.
In practice, $\Gamma^A_{iajb}$ and $\Gamma^B_{iajb}$ are constructed iteratively using the the bosonic density operator $\hat{\Gamma}^B_{iajb} = \frac{1}{2} \left( \hat{b}^\dagger_{ia} \hat{b}^\dagger_{jb} + \hat{b}_{ia} \hat{b}_{jb} \right)$ and $\Gamma^A_{iajb} = \hat{b}^\dagger_{ia} \hat{b}_{jb} $, as
\begin{align}
    \bar{\Gamma}^{A,B}(n+1) = \left[\bar{\Gamma}(n),\hat{\sigma}\right],
\end{align}
until the norm $\frac{||\bar{\Gamma}^{A,B}(n)||}{n!}<10^{-9}$. This procedure is analog to the construction of $\bar{H}$. Details are presented in Appendix A of Ref.~\citenum{tolle2024ab}.
\\
The quasiparticle density matrices are obtained analogously from the derivative of $E^\mathrm{QP}_\nu$\

\begin{align}
    \gamma^\mathrm{QP}_{\nu,pq} = \frac{\partial E^\mathrm{QP}_\nu }{\partial F_{pq}},
\end{align}
and
\begin{align}
    \Gamma^\mathrm{QP}_{\nu,pqrs} = \frac{\partial E^\mathrm{QP}_\nu }{\partial V_{pqrs}}.
\end{align}
The equations are 
\begin{align}
    \gamma^\mathrm{QP}_{\nu,ij} &=  R_{\nu,i} R_{\nu,j} + \sum_{I} R_{\nu,iI} R_{\nu,jI}  \nonumber \\
    &+ \sum_{ka}  \tilde{R}_{\nu,k(ja)}\tilde{R}_{\nu,k(ia)} - \sum_{ba}  \tilde{R}_{\nu,(ib)a}\tilde{R}_{\nu,(jb)a}  \nonumber \\
    \gamma^\mathrm{QP}_{\nu,ab} &=R_{\nu,a} R_{\nu,b} + \sum_{I}  R_{\nu,Ia} R_{\nu,Ib}  \nonumber \\
    &+\sum_{ac} \tilde{R}_{\nu,(ia)c} \tilde{R}_{\nu,(ib)c} - \sum_{kj} \tilde{R}_{\nu,k(ja)} \tilde{R}_{\nu,k(jb)} \nonumber \\
    \gamma^\mathrm{QP}_{\nu,ia} &= R_{\nu,i} R_{\nu,a} \nonumber \\
    \gamma^\mathrm{QP}_{\nu,ai} &= R_{\nu,a} R_{\nu,i},
\end{align} 
with $\tilde{R}_{\nu,Ip}$ denoting 
\begin{align}
    \tilde{R}_{\nu,Ip} = \sum_{J}R_{\nu,Jp}\left[e^{\mathbf{t}}\right]_{JI}.
\end{align}
In practice, $e^{\mathbf{t}}$ is evaluated by Taylor expansion until the norm of the n-th term $\frac{\mathbf{||T||}}{n!}<10^{-12}$.
For $\Gamma^\mathrm{QP}_{\nu,pqrs}$, one finds
\begin{align}
   \Gamma^\mathrm{QP}_{\nu,iajb} &= 4 \left( \sum_{c} \tilde{R}_{\nu,(ia)c} \tilde{R}_{\nu,(jb)c} -  \sum_{k} \tilde{R}_{\nu,k(ia)} \tilde{R}_{\nu,k(jb)} \right.\nonumber \\
   &\left. +  R_{\nu,i} \tilde{R}_{\nu,(jb)a} + R_{\nu,a}\tilde{R}_{\nu,i(jb)} \right)\nonumber \\
   \Gamma^\mathrm{QP}_{\nu,ijkc} &= 2R_{\nu,i} \tilde{R}_{\nu,j(kc)} \nonumber \\
   \Gamma^\mathrm{QP}_{\nu,kcij} &= 2\tilde{R}_{\nu,j(kc)} R_{\nu,i} \nonumber \\
   \Gamma^\mathrm{QP}_{\nu,abkc} &= 2R_{\nu,a} \tilde{R}_{\nu,(kc)b} \nonumber \\
   \Gamma^\mathrm{QP}_{\nu,kcab} &= 2\tilde{R}_{\nu,(kc)b} R_{\nu,a}.
    \label{eq:QP2RM}
\end{align}
In the case of the TDA, the prefactor for the first term in Eq.~(\ref{eq:QP2RM}) changes to two.
\\
The density matrices associated with the $\mathbf{Z}$-Lagrange Multiplier requires the determination of these multipliers.
They are determined from 
\begin{align}
    \frac{\partial L^\mathrm{QP}_{\nu}(\mathbf{Z}, \pmb{\lambda}, \mathbf{M}, \mathbf{C}, \mathbf{t})}{\partial t_{KL}} &\overset{!}{=} 0.
\end{align}
$\mathbf{Z}$ is determined from solving the following system of linear equations
\begin{align}
     \frac{\partial E^\mathrm{drUCCD}_\nu}{\partial t_{KL}} + \frac{\partial E^\mathrm{QP}_\nu}{\partial t_{KL}} + \sum_{IJ} Z_{IJ} \frac{\partial J_{IJ}}{\partial t_{KL}} = 0.
\end{align}
Because $\frac{\partial E^\mathrm{drUCCD}_\nu}{\partial t_{\nu,\mu}} = 0$, this simplifies to
\begin{align}
    \frac{\partial E^\mathrm{QP}_\nu}{\partial t_{KL}} + \sum_{IJ} Z_{IJ} \frac{\partial J_{IJ}}{\partial t_{KL}} = 0.
    \label{eq:ZVectorEquation}
\end{align}
For the first term $\frac{\partial E^\mathrm{QP}_\nu}{\partial t_{KL}}$ (only showing non-zero terms), one finds 
\begin{align}
    \frac{\partial E^\mathrm{QP}_\nu}{\partial t_{KL}} &= \frac{\partial}{\partial t_{KL}} \left( \sum_{ijI} R_{\nu,i} W^I_{ij} R_{\nu,jI} + \sum_{ijI} R_{\nu,iI} W^I_{ij} R_{\nu,j} \right. \nonumber \\
    &+ \left. \sum_{abI} R_{\nu,a} W^I_{ab} R_{\nu,Ib} + \sum_{abI} R_{\nu,bI} W^I_{ab} R_{\nu,a} \right. \nonumber \\ 
    &+ \left. \sum_{iaI} R_{\nu,i} W^I_{ia} R_{\nu,Ia} + \sum_{iaI} R_{\nu,Ia} W^I_{ai} R_{\nu,i} \right. \nonumber \\
    &+ \left.\sum_{iaI} R_{\nu,a} W^I_{ai} R_{\nu,iI} + \sum_{iaI} R_{\nu,iI} W^I_{ia} R_{\nu,a} \right. \nonumber \\
    &- \left. \sum_{iIJ} R_{\nu,iI} \bar{A}_{IJ} R_{\nu,iJ} + \sum_{aIJ} R_{\nu,Ia} \bar{A}_{IJ} R_{\nu,Ja}  \right),
\end{align}
with 
\begin{align}
    W^I_{pq} = \sum_{J} V^{J}_{pq} \left[ e^\mathbf{t} \right]_{JI},
\end{align}
and 
\begin{align}
    \bar{A}_{IJ} = \sum_{KL}\left[ e^\mathbf{t} \right]_{IK} \left[\mathbf{A} + \mathbf{B}\right]_{KL} \left[ e^\mathbf{t} \right]_{LI}.
\end{align}
The derivative requires the differentiation of $\left[ e^\mathbf{t} \right]_{IJ}$
\begin{align}
    \frac{\partial \left[ e^\mathbf{t} \right]_{IJ} }{\partial t_{KL}} = \sum^M_{n=0} \frac{1}{\left(n+1\right)!} \sum^{n}_{k=0} \left[ \mathbf{t}^{n-k}\right]_{IK} \left[\mathbf{t}^{k}\right]_{LJ}.
\end{align}
Throughout this manuscript, the derivative is approximated with $M=5$, which has been found to be sufficient (compare Tabs.~\ref{tab:EOMIPGRAD},\ref{tab:EOMEAGRAD}).
Similarly the gradient for
\begin{align}
    \frac{\partial R_{IJ}}{\partial t_{KL}} &= \frac{\partial }{\partial t_{KL}} \left( \bar{\mathbf{B}} + \bar{\mathbf{B}}^T \right)_{IJ} 
\end{align}
is obtained, after rewriting $\bar{\mathbf{B}}$ as 
\begin{align}
    \bar{\mathbf{B}} &= \sum_{KL}\left[ e^\mathbf{t} \right]_{IK} \left[\mathbf{A} + \mathbf{B}\right]_{KL} \left[ e^\mathbf{t} \right]_{LI} \nonumber \\ 
    &+ \sum_{KL}\left[ e^{-\mathbf{t}} \right]_{IK} \left[\mathbf{B} - \mathbf{A}\right]_{KL} \left[ e^{-\mathbf{t}} \right]_{LI}.
\end{align}
\\
With $\mathbf{Z}$  [Eq.~(\ref{eq:ZVectorEquation})], the density matrices   $\gamma^\mathrm{Z}_{pq}$ and $\Gamma^\mathrm{Z}_{pqrs}$ are obtained by differentiation
\begin{align}
   \gamma^\mathrm{Z}_{pq} = \frac{\partial \sum_{iajb} Z_{ia,jb} J_{ia,jb}}{\partial F_{pq}} \nonumber \\
   \Gamma^\mathrm{Z}_{pqrs} = \frac{\partial \sum_{iajb} Z_{ia,jb} J_{ia,jb}}{\partial V_{pqrs}}.
\end{align}
\section{Appendix B: Electron addition and removal processes}
\label{sec:AppendixB}
\begin{table}[h]
    \caption{Experimental adiabiatic ionization potentials (aIPs) and electron affinities (aEAs) for various small molecular systems in eV. Reference values are taken from Ref.~\citenum{curtiss1991gaussian}.}
    \begin{tabular}{l r}
        \hline 
        Process &  Exp.\\
        \hline
        C$_2$H$_2$$\rightarrow$C$_2$H$_2^{+}$ & 11.40 \\
        C$_2$H$_4$$\rightarrow$C$_2$H$_4^{+}$ & 10.51\\
        CO $\rightarrow$CO$^{+}$ & 14.01\\
        Cl$_2$$\rightarrow$Cl$_2^{+}$ & 11.50\\
        ClF $\rightarrow$ClF$^{+}$ & 12.66\\
        ClH $\rightarrow$ClH$^{+}$ & 12.75\\
        FH $\rightarrow$FH$^{+}$ & 16.04\\
        H$_2$O $\rightarrow$H$_2$O$^{+}$ & 12.62\\
        N$_2$ $\rightarrow$N$_2^{+}$ & 16.70\\
        P$_2$ $\rightarrow$P$_2^{+}$ & 10.53\\
        PH$_3$ $\rightarrow$PH$_3^{+}$ & 9.87\\
        NH$_3$ $\rightarrow$NH$_3^{+}$ & 10.18\\
        SiH$_4$$\rightarrow$SiH$^+_4$& 11.00\\
        \hline 
        Cl$_2$$\rightarrow$Cl$_2^{-}$ & 2.39\\
        SiH$_2$$\rightarrow$SiH$_2^{-}$ & 1.12\\
        \hline
    \end{tabular}
    \label{tab:Processes}
\end{table}
\bibliography{literatur}
\end{document}